\documentstyle[prc,aps,preprint]{revtex}
\begin{document}
\draft

\title{Momentum Distributions of Particles from Three--Body Halo
Fragmentation: Final State Interactions}
\author{E.~Garrido, D.V.~Fedorov and A.S.~Jensen \\
Institute of Physics and Astronomy, \\
Aarhus University, DK-8000 Aarhus C, Denmark}
\date{\today}

\maketitle

\begin{abstract}
Momentum distributions of particles from nuclear break-up of fast
three-body halos are calculated consistently, and applied
to $^{11}$Li. The same two-body interactions between the three particles 
are used to calculate the ground state structure and 
the final state of the reaction processes. We reproduce the available
momentum distributions from $^{11}$Li fragmentation, together with the
size and energy of $^{11}$Li, with a neutron-core relative state 
containing a $p$-state admixture of 20\%-30\%. The available fragmentation 
data strongly suggest an $s$-state in $^{10}$Li at about 50 keV, and
indicate a $p$-state around 500 keV.


\end{abstract}

\newpage

{\it Introduction.} The existence of halo nuclei is by now well
established as weakly bound and spatially extended systems
\cite{hansen1}. The dominant features of these unusual nuclei are most
easily described by few-body models. The many degrees of freedom are
then divided into the approximately frozen (core) and the active
(halo) degrees of freedom. Substantial efforts have been allocated to
investigations of three-body halos (one core and two halo particles)
and in particular to bound three-body systems where all two-particle
subsystems are unbound \cite{dima2}. The nuclear prototypes of these
so-called Borromean systems are $^{6}$He ($^{4}$He+n+n) and $^{11}$Li
($^{9}$Li+n+n), both thoroughly discussed in a general framework in
\cite{zhukov1}.

The most detailed source of experimental information about the
structure of these nuclei is the momentum distributions of the
``particles'' resulting from fragmentation reactions
\cite{anne,zinser,nilsson,humbert}.  One major problem in the
interpretation of such measurements is the inherent mixture of effects
from the original structure and the reaction mechanism \cite{carlos}. 
Both should therefore be consistently incorporated in model calculations.

The simplest model assumes the sudden approximation, where one of the
particles instantaneously is removed from the three-body system while
the other two particles remain completely undisturbed.  Clearly this
can only be justified for reaction times much shorter than the
characteristic time for the motion of the three particles in the
system. Since halo nuclei are weakly bound 
and the beam energy is very high this requirement is exceedingly well 
fulfilled. The observed momentum distributions therefore seems to provide 
direct information about the three-body wave function \cite{thompson}. 
However, the interactions between the remaining two particles are often
essential especially when low lying resonances are present
\cite{zinser,barranco,korsheninnikov}. In principle, processes in which
the three particles move along together after the fragmentation are
also possible. However they occur through Coulomb dissociation, requiring 
a heavy target or by using of a low energy beam. 
Since we consider a light target and a high energy beam this kind
of reactions is then out of our model, and they are not considered in
this letter.

The final state two-body interactions (FSI) are active in the last part of
the reaction process and they determine simultaneously the three-body
structure of the initial halo nucleus. A consistent treatment is
necessary to allow reliable interpretations of this major source of 
detailed experimental information. In this letter we report on such model
calculations, where we combine an accurate three-body description with
an equally accurate computation of momentum distributions. After a
general discussion we shall present detailed calculations of momentum
distributions of particles from fragmentation reactions of $^{11}$Li
when FSI are included. 

{\it Method.} After the collision in which one particle is suddenly
removed, the probability of finding the remaining two particles 
with relative momentum $\mbox{\bf k}_x$ and total momentum
$\mbox{\bf k}_y$ relative to the center of mass of the three-body
system is proportional to the overlap
\begin{equation}
\Psi(\mbox{\bf k}_x, \mbox{\bf k}_y) = 
\langle e^{i \mbox{\scriptsize\bf k}_y \cdot \mbox{\scriptsize\bf y} }
        e^{i \mbox{\scriptsize\bf k}_x \cdot \mbox{\scriptsize\bf x} }  |
        \Psi(\mbox{\bf x}, \mbox{\bf y}) \rangle,
\label{eq1}
\end{equation}
where $\Psi(\mbox{\bf x}, \mbox{\bf y})$ is the three-body wave 
function. The coordinates $\mbox{\bf x}$ and $\mbox{\bf y}$ are the
usual Jacobi coordinates \cite{dima2,zhukov1} where 
$\mbox{\bf x}$ is drawn between
the two particles surviving after the fragmentation.
Without FSI the momentum distribution or the
differential cross section is then proportional to the square of the
Fourier transform of the three-body wave function.

Inclusion of FSI now amounts to
substituting the plane wave $e^{i \mbox{\scriptsize\bf k}_x \cdot
\mbox{\scriptsize\bf x} }$
in eq.(\ref{eq1}) by the appropriate {\it distorted}
two-body wave function $w$. The momentum distribution is then given by
\begin{equation}
\frac{d^6 \,\sigma}{d\,\mbox{\bf k}_x d\,\mbox{\bf k}_y} \propto
\sum_M \sum_{s_x, \sigma_x}
|\langle e^{i \mbox{\scriptsize\bf k}_y \cdot \mbox{\scriptsize\bf y} }
        w^{s_x \sigma_x}(\mbox{\bf k}_x, \mbox{\bf x} )
|
        \Psi^{JM}(\mbox{\bf x}, \mbox{\bf y}) \rangle|^2
\label{eq3}
\end{equation} 
where $J$ is the total spin of the halo nucleus, and
$s_x$ and $\sigma_x$ are the spin of the two-body final
state and its projection. The
summations in eq.(\ref{eq3}) arise from the average over
initial states ($M$) and the sum over final states ($s_x$ and $\sigma_x$). 

In our calculation the three-body wave function is obtained by
solving the Faddeev equations in coordinate space, where the
nucleon-nucleon potential is fitted to low energy $s$ and $p$-wave
nucleon-nucleon scattering data, and the neutron-core potential 
is adjusted to give the proper binding energy and mean square radius 
of the three-body system \cite{dima4,dima5}. 

The partial wave expansion of the two-body final
state wave function is written as \cite{newton} 
\begin{equation}
w^{s \sigma}(\mbox{\bf k}, \mbox{\bf x}) =
\sqrt{\frac{2}{\pi}} \frac{1}{k x} \sum_{j, \ell, m}
u_{\ell s}^{j}(k,x) {\cal Y}_{j \ell s}^{m^{\mbox{\normalsize $\ast$}}}
(\Omega_x)
\sum_{m_\ell=-\ell}^{\ell} \langle \ell s ; m_\ell \sigma|j m\rangle
i^\ell Y_{\ell m_\ell}(\Omega_k) ,
\end{equation}
where
the radial functions $u_{\ell s}^{j}(k,x)$ are obtained by solving
the Schr\"{o}dinger equation with the appropriate two-body
potential. Finally, we calculate the expression in eq.(\ref{eq3}) and
subsequently integrate over the unobserved quantities to obtain the
measured momentum distributions.  The details of the formalism will be
presented elsewhere.

{\it General results.} We reveal the general results of the procedure
described above by using the three-body halo $^{11}$Li ($^{9}$Li+n+n) 
as an example, and neglecting the spin of the core. The neutron-neutron 
interaction is fitted as previously indicated, and 
the neutron-core $s$ and $p$-wave potentials are varied to 
study the dependence of the momentum distribution on the positions 
of the resonances in $^{10}$Li, as well as the effect of the inclusion of 
FSI. The binding energy and mean square radius of $^{11}$Li is kept fixed to
the experimental values.

In the upper part of Fig.1 we show the resulting neutron momentum
distributions after one of the neutrons is removed by the target for
three different energies of the $s_{1/2}$ virtual state (100, 200 and
300 keV). The upper set of curves is obtained without FSI. They show 
a little variation,
since they depend mainly upon the binding energy and r.m.s. radius of
$^{11}$Li. The final state interactions substantially reduce the widths,
the stronger the lower the virtual state is. In table I we summarize the 
full width at half maximum (FWHM) for the curves in Fig.1. Here we also 
give the FWHM of the core one dimensional momentum distributions, which 
are much less influenced by FSI due to the larger mass.

The lower part of Fig.1 shows the same kind of calculation for core
break-up reactions, where the core is destroyed during the
interaction, keeping the validity of the sudden approximation. 
The interest in this kind of reactions is that now the
final state interaction (neutron-neutron interaction) is well known,
reducing the uncertainties involved in the calculation. We first
observe that variation of the position of the $s_{1/2}$ virtual state 
only produces a small change in the momentum distributions. That is 
expected, since the final state neutron-neutron interaction is independent
of the properties of $^{10}$Li (n+core). However when comparing with
the results without FSI, as shown by the upper set of curves, we still 
observe a noticeable variation (note that the curves without FSI are 
identical to the corresponding curves in the upper part of the figure).
Now the FWHM of the distribution is around 40 MeV/c, broader
than the distributions obtained when one of the neutrons is removed,
but narrower than the calculation without FSI.
As seen from table I, the agreement with the experimental data is
quite good, especially for low energies of the $s$ virtual state. This
is in complete agreement with the low lying $s$ virtual state ($\sim
50$ keV) recently suggested in \cite{zinser} along with a $p$-state
resonance near 0.5 MeV.

It should be mentioned that a narrow 
momentum distribution can also be obtained with a low lying
$p$-resonance. However in this case the lowest $s$-state must be shifted
towards much higher energies to reproduce the correct binding energy 
of $^{11}$Li.

{\it Realistic calculations for $^{11}$Li.} Up to now, we have neglected the
spin dependence of the neutron-core interaction or equivalently
assumed that the spin of the core is zero. Of course, this is not
realistic (the spin of $^9$Li is 3/2), and the neutron-core
interaction should include a spin dependence, splitting the two
possible $s$-states of $^{10}$Li with total angular momentum 1 and 2.
To do this, we have in the neutron-core potential included one term
proportional to $\mbox{\bf s}_n \cdot \mbox{\bf s}_c$, where
$\mbox{\bf s}_n$ is the spin of the neutron and $\mbox{\bf s}_c$ the
spin of the core \cite{dima5}. For simplicity, the spin splitting term
has been introduced only in the $s$ wave.

It is now possible to place an $s$-state at 50 keV and the lowest
$p$-resonance at 500 keV, as suggested by the analyses in \cite{zinser}.
Simultaneously we are able to vary the content of $s$ and $p$-waves in
the neutron-core subsystem of $^{11}$Li. Such realistic
calculations can be compared directly with the experimental data.
We then place the lowest $p$-resonance at 500 keV, and the
lowest $s$-state with total angular momentum 2 at 50 keV. Then we use
the spin-orbit potential in the neutron-core interaction to vary the
total $p$-state content in the $^{11}$Li wave function, and finally we
adjust the energy of the second $s$-state (with total angular momentum
1) to recover the correct binding energy and mean square radius in the
$^{11}$Li. This procedure determines completely the low lying
resonance structure of $^{10}$Li. Very similar results are obtained by
placing the $s$-state with angular momentum 1 at 50 keV, and instead
adjusting the energy of the $s$-state with angular momentum 2.

In Fig.2 we show the one dimensional transverse momentum distribution
of $^9$Li fragments. The experimental data \cite{humbert} correspond
to a reaction at 280 MeV/u in a C target. The $^{11}$Li wave function
contains around 26\% of $p$-wave in the neutron-core subsystem, but the
result is almost independent of the $p$-state admixture. 
The results of the computation with (solid line) and without (dashed line) 
FSI are shown in the figure. The curves have been 
convoluted with the experimental beam profile \cite{humbert}. 
As expected, the effect of FSI is rather small in
the central region. For momenta larger than around 75 MeV/c the
neutrons are inside the core, and our three-body model is not valid anymore. 
Also, in the high momentum region other effects, like diffraction processes, 
should be considered. The good agreement in the central region shows that our 
$^{11}$Li wave function is accurate enough to describe the process. 

To evaluate the effect of FSI on the momentum
distributions we investigate neutron distributions, much more sensitive
to them. The results are shown in Fig.3 for the two dimensional neutron
momentum distribution, both for neutron removal process (upper part) and 
for core break-up reactions (lower part). The appropriate FSI
are included, i.e.\ neutron-core in the first case and
neutron-neutron in the second case.  In both cases the experimental data
correspond to reactions of $^{11}$Li at 280 MeV/u with a C
target\cite{zinser,nilsson}.  Several $^{11}$Li wave functions with
different content of $s$ and $p$-waves in the neutron-core channel
have been used. In particular, going from the narrower to the broader
distributions, the curves correspond to calculations with 4, 18,
and 35\% of $p$-wave. In both parts of the figure, FSI
are essential to recover the observed behaviour of the
distributions.  

The core break-up process (lower part) is almost insensitive to the
structure of $^{11}$Li and the peculiarities of the neutron-core
interaction. This is due to the fact that the final state interactions
between the two neutrons are independent of the neutron-core
potential. The nice agreement with the experiment strongly
supports the method and the model. 
In the upper part of Fig.3 the two dimensional neutron momentum distribution 
for neutron removal
reactions are shown. Now we see that a $^{11}$Li wave function with a
small content of $p$-wave clearly underestimates the width of the
momentum distribution, while a wave function with more than 35\% of
$p$-wave in the neutron-$^9$Li subsystem overestimate the
width. The best results are obtained when a $p$-wave admixture of
around 26\% is used. The data in the upper part of Fig.3 also support
the choice of a 500 keV $p$-state, since this energy corresponds to the
bump observed around $p_r \simeq 30$ MeV/c.

{\it Summary and conclusions.} We have calculated momentum
distributions of fragments from high-energy nuclear break-up reactions
of three-body Borromean halo nuclei. The two-body interaction in the
three-body description of the halo nucleus is identical to the final
state interaction between the two particles remaining after the target
has removed the third particle. We maintain this consistency in the
description. The final state interaction may significantly reduce the
width of the momentum distribution of the light particle. The size of
this reduction strongly depends on the resonance structure of the
remaining two-body system.

After extraction of the general properties, we applied the model to
fragmentation of the three-body halo nucleus $^{11}$Li. First, we
compared the core-momentum distribution after neutron removal with
experimental data. The results are only marginally sensitive to final state
interactions and the sudden approximation only recovers the ground state 
wave function in
momentum space. The good agreement with measurements therefore
demonstrates the validity of our $^{11}$Li three-body model and the
reaction mechanism assumed. 

Secondly, we computed two dimensional neutron distributions after core
break-up. They are rather insensitive to the characteristics of the
neutron-core interaction which only enters through the properties of
the ground state wave function. However the final state
neutron-neutron interaction is essential to reproduce the measured
data. This demonstrates the validity of our method of including final
state interactions.

Thirdly, we computed two dimensional neutron distributions after one
neutron removal. They depend crucially on the neutron-core interaction
which now enters both through the properties of the ground state
wave function and through the final state interaction. The structure of
$^{11}$Li must then contain a $p$-state admixture of about 26\% in the
relative neutron-core system. The same interaction determines the
structure of $^{10}$Li, where a $p$-state resonance at about 500~keV
produces the small shoulder at about 30~MeV/c in the experimental
distribution. The statistical average of the two spin splitted $s$
virtual states must then be at about 700~keV in order to reproduce
the binding energy of $^{11}$Li. Furthermore, the narrow momentum
distribution can only be reproduced with the lowest $s$ virtual state
in $^{10}$Li at about 50~keV. 

{\bf Acknowledgments} One of us (E.G.) acknowledges support from the
European Union through the Human Capital and Mobility program contract
nr. ERBCHBGCT930320.

\begin{table}
\begin{tabular}{cccccc}
     & \multicolumn{3}{c}{neutron} & \multicolumn{2}{c}{core} \\ \cline{2-6}
  &  no FSI & FSI$^{\mbox{\scriptsize (n-c)}}$  
            & FSI$^{\mbox{\scriptsize (n-n)}}$ & no FSI 
            & FSI$^{\mbox{\scriptsize (n-c)}}$ \\
\cline{2-6}
$E_{s_{1/2}}=100$ keV &   51.1  & 28.3 &       38.4          & 68.5   & 55.1 \\
$E_{s_{1/2}}=200$ keV &   53.2  & 33.9 &       39.0          & 69.8   & 58.0 \\
$E_{s_{1/2}}=300$ keV &   55.8  & 39.9 &       39.7          & 71.4   & 60.4 \\
Experimental  &         & 25-35$^{\mbox{\scriptsize (a)}}$& 
                                  $43\pm 3^{\mbox{\scriptsize (b)}}$&     &
                                  $49\pm 3^{\mbox{\scriptsize (c)}}$
\end{tabular}
\vspace{1cm}

\caption{
Comparison of the full width at half maximum in MeV/c of the
longitudinal momentum distributions from $^{11}$Li fragmentation for
different positions of the $s_{1/2}$ virtual state. Columns 2 to 4
refer to neutron momentum distributions, while 5 and 6 refer to core
momentum distributions. (n-c) FSI between $^9$Li
and the neutron have been included (neutron removal process). (n-n)
FSI between the two neutrons have been included
(core break-up process). (a) See, for instance, \protect\cite{anne}. (b)
Data from \protect\cite{nilsson}. (c) Data from
\protect\cite{humbert}.  }
\end{table}

\newpage
\begin{center}
{\Large\bf Figure Captions}
\end{center}

{\bf Fig.1.-} Upper part: The neutron momentum distributions
from neutron removal process in $^{11}$Li fragmentation. $p_x$ is the
$x$-cartesian component of the momentum of the neutron relative to the
center of mass of the three-body system. The
$s_{1/2}$ virtual state is placed at 100 keV (solid line), 200 keV
(dashed line), and 300 keV (long dashed line), respectively. The upper
set of curves corresponds to calculations without FSI, while FSI are 
included in the lower set. Lower part: The same as before for core break-up
reactions.

{\bf Fig.2.-} Transverse core momentum distribution after neutron
removal from $^{11}$Li. $p_x$ is the $x$-cartesian component of the 
momentum of the core relative to the center of mass of the three-body 
system. The solid (dashed) line is the calculation
with (without) FSI. Data from \protect\cite{humbert}.

{\bf Fig.3.-} Radial neutron momentum distribution for neutron 
removal (upper part) and core break-up (lower part) reactions of
$^{11}$Li. $p_r=\protect\sqrt{p_x^2+p_y^2}$ is the cylindrical radial
component of the momentum of the neutron relative to the center of mass
of the three-body system. Results with (solid line) and without 
(dashed line) FSI are presented. 
The $p$ wave content in the neutron-$^9$Li subsystem (from narrower to 
broader distributions) is 4, 18, and 35\%. Experimental data are taken 
from \protect\cite{nilsson} and \protect\cite{zinser}.


\begin{thebibliography}{99}
\bibitem{hansen1} P.G. Hansen, A.S. Jensen and B. Jonson, 
to be published in Ann. Rev. Nucl. Part. Sci. {\bf 45} (1995).  
\bibitem{dima2} D.V. Fedorov, A.S. Jensen, and K. Riisager, Phys. Rev.  
                    {\bf C49}, 201 (1994).
\bibitem{zhukov1} M.V. Zhukov, B.V. Danilin, D.V. Fedorov, J.M. Bang,
     I.S. Thompson, and J.S. Vaagen, Phys. Rep. {\bf 231}, 151 (1993).
\bibitem{anne} R. Anne {\it et al.}, Phys. Lett. {\bf B250}, 19 (1990).
\bibitem{zinser} M. Zinser {\it et al.}, Phys. Rev. Lett. 
                    {\bf 75}, 1719 (1995).
\bibitem{nilsson} T. Nilsson {\it et al.}, Europhys. Lett. {\bf 30}, 19 (1995).
\bibitem{humbert} F. Humbert {\it et al.}, Phys. Lett. {\bf B347}, 198 (1995).
\bibitem{carlos} C.A. Bertulani, L.F. Canto, and M.S. Hussein,
Phys. Rep. {\bf 226}, 281 (1993).
\bibitem{thompson}  I.S. Thompson and M.V. Zhukov, Phys. Rev. 
    {\bf C49}, 1904 (1994).
\bibitem{barranco} F. Barranco, E. Vigezzi, and R.A. Broglia, Phys.
                Lett. {\bf B319} 387 (1993)
\bibitem{korsheninnikov} A.A. Korsheninnikov and T. Kobayashi, Nucl.
                Phys. {\bf A567}, 97 (1994).
\bibitem{dima4} D.V. Fedorov, A.S. Jensen, and K. Riisager, Phys. Rev.  
                    {\bf C50}, 2372 (1994).
\bibitem{dima5} D.V. Fedorov, E. Garrido , and A.S. Jensen, Phys. Rev.  
                     {\bf C51}, 3052 (1995).
\bibitem{newton} R.G. Newton, Scattering Theory of Waves and Particles,
(Springer-Verlag, N.Y., 1982), Second Edition, p.444.

\end{thebibliography}
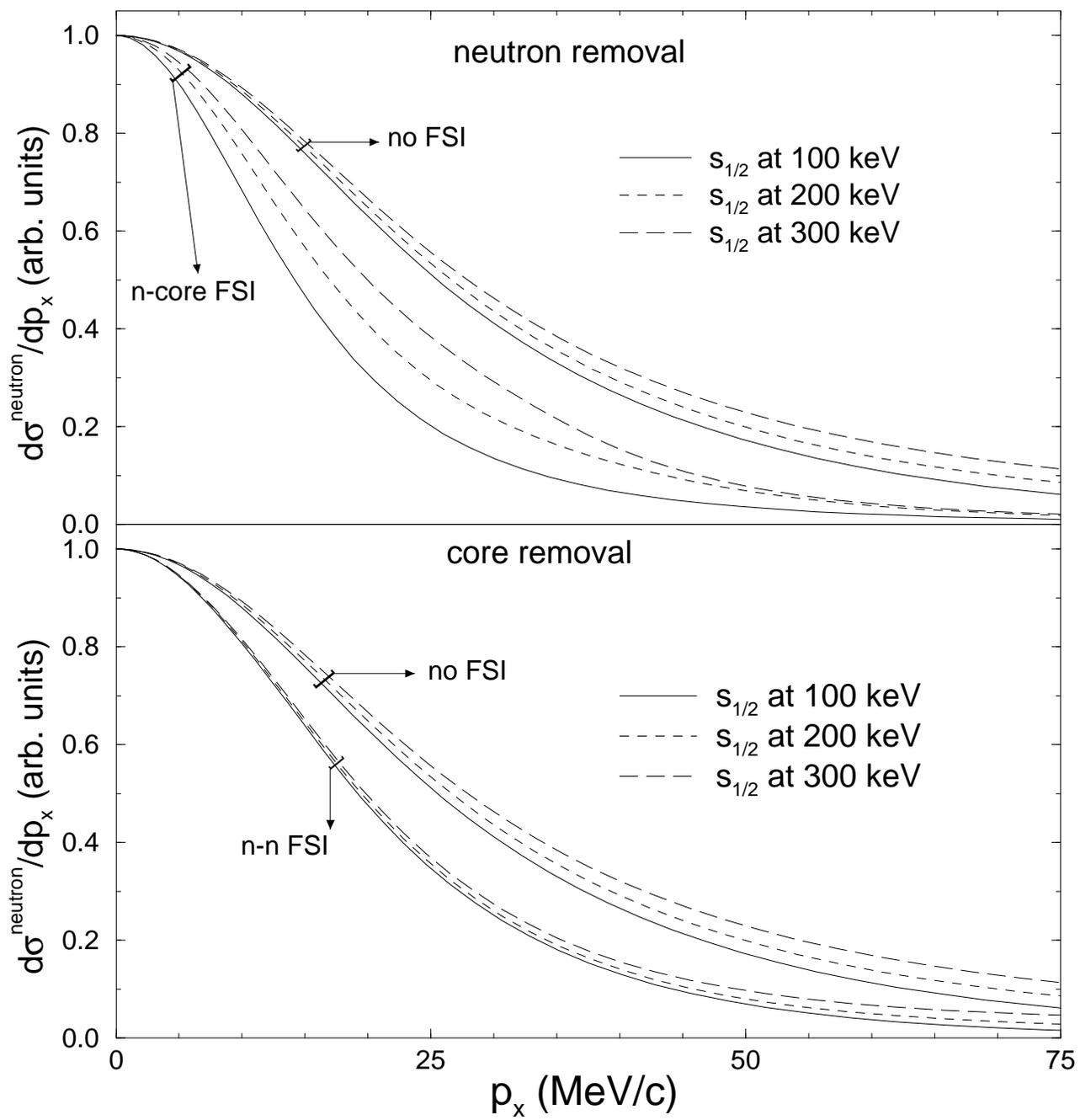
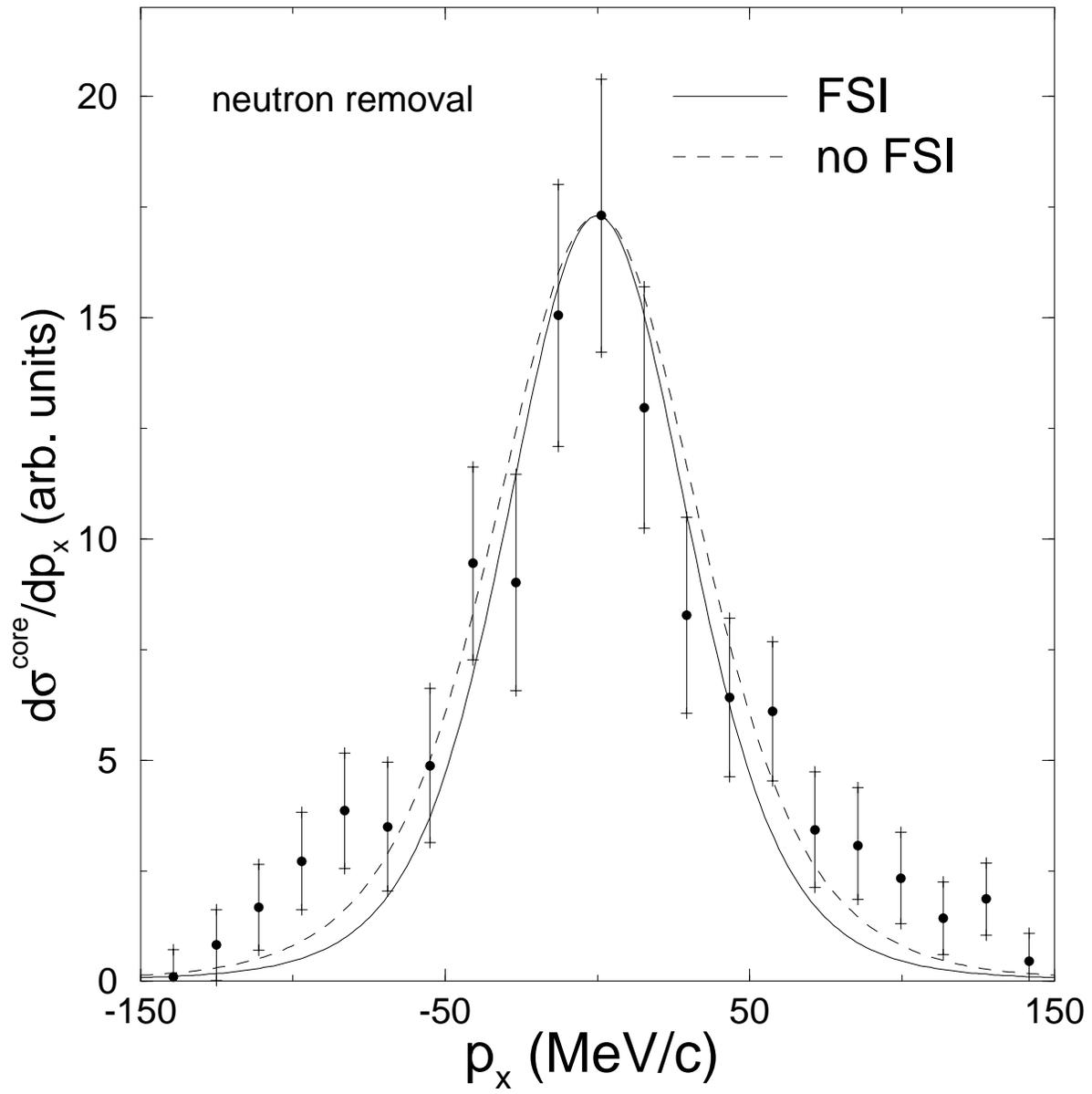
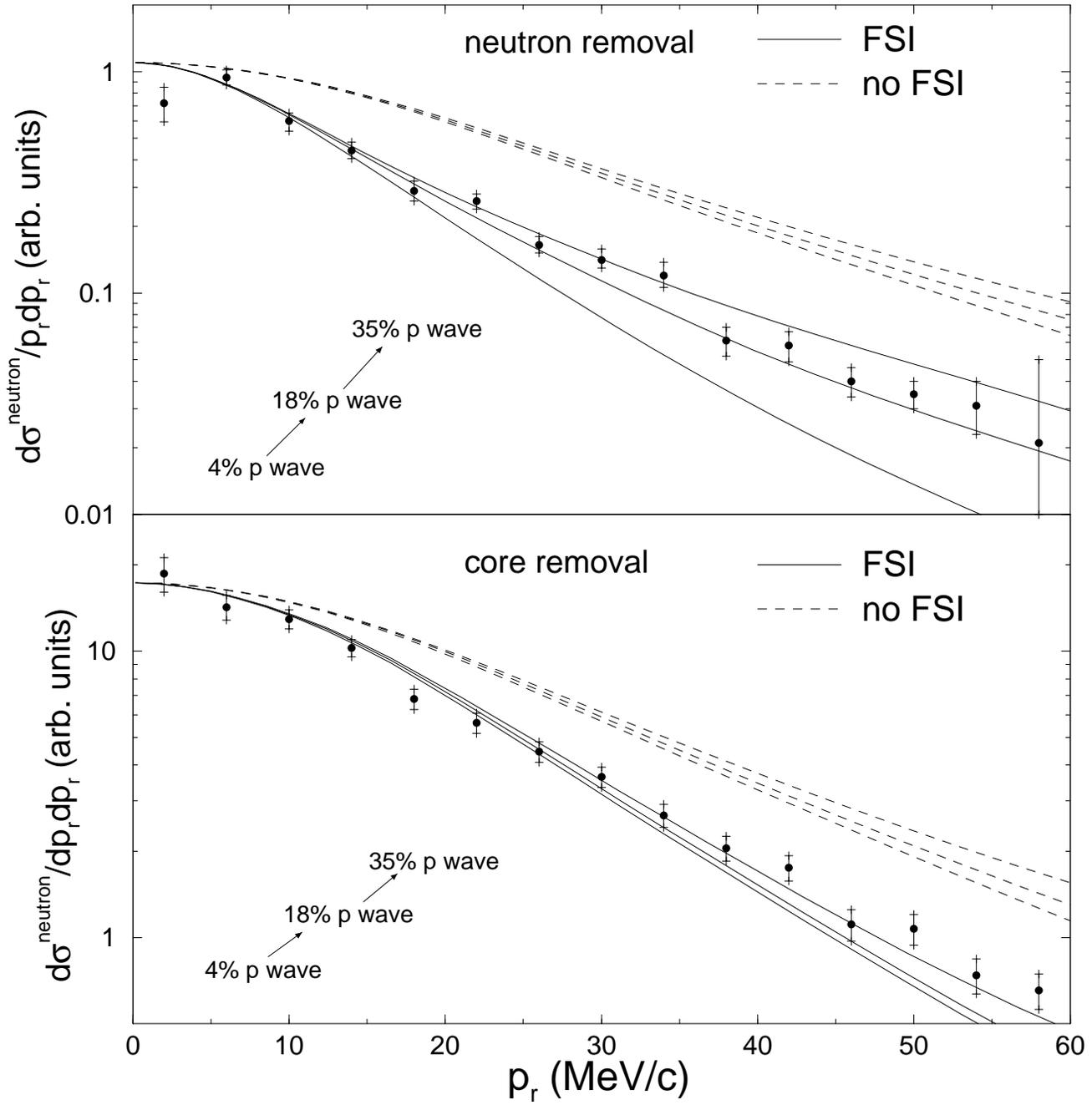
\end{document}